Title : **Left- *vs* right-handed badminton slice shots: opposite shuttlecock spinning and Magnus effect**

Titre : Slices de gaucher vs droitier au badminton : rotation de volant inverse et effet Magnus

Eric Collet,[a,b,c,]*

eric.collet@univ-rennes.fr

[a] Univ Rennes, CNRS, IPR (Institut de Physique de Rennes) - UMR 6251, F-35000 Rennes, France

[b] DYNACOM IRL2015 University of Tokyo - CNRS - UR1, Department of Chemistry, 7-3-1 Hongo, Tokyo 113-0033 – Japan

[c] Institut universitaire de France (IUF)

Eric Collet : https://ipr.univ-rennes1.fr/interlocuteurs/eric-collet

- Abstract

The chiral nature of a badminton shuttlecock is responsible for its anti-clockwise spinning as it naturally propagates through the air. This induces a dissymmetry between left-handed and right-handed players and the resulting trajectories of the shuttlecock, which were captured in real condition on the badminton court and in slow-motion at 3700 fps. The videos clearly evidence this dissymmetry as slice shots performed by righties accelerate the natural anti-clockwise spinning, while the one performed by lefties induces a clockwise to anti-clockwise spinning, making trajectories of shuttlecocks different. The slow-motion videos also caught a brief Magnus effect, often neglected in badminton, lifting up the shuttlecock for both lefties and righties and affecting the effectiveness of the slice shot.

- Key words

Physics of badminton, Magnus effect, symmetry, chirality, badminton trajectory, time-resolved

- Résumé

La nature chirale du volant de badminton est responsable de sa rotation anti-horaire lorsqu'il se propage naturellement dans l'air. Cela induit une dissymétrie entre joueurs gauchers et droitiers et les trajectoires du volant qui en résultent, et qui ont été capturées en condition réelle sur le terrain de badminton en slow motion à 3700 ips. Les vidéos mettent clairement en évidence cette dissymétrie puisque les slices effectués par les droitiers accélèrent la rotation anti-horaire naturelle, tandis que ceux effectués par les gauchers induisent une rotation d'horaire à anti-horaire, rendant les trajectoires de volants différentes. Les vidéos ont également mis en évidence un bref effet Magnus, souvent négligé au badminton, soulevant le volant pour les gauchers comme les droitiers et affectant l'efficacité du slice.

## 1. Introduction

The Badminton World Federation reported that badminton is one of the most popular sports, played by over 339 million people worldwide. About 10% of them are left-handed players. While in most sports there is no geometrical difference between left-handed and right-handed players, this is not true for badminton. This dissymmetry comes from the chiral nature of feather shuttlecocks, composed of an array of diverging stems, the ends of which are at the convergent end of the skirt, joined together in an end ring. Figure 1 shows the nomenclature of feather shuttlecocks. Its global symmetry corresponds to the point group $C_{16}$, which contains 2-, 4-, 8- and 16-fold rotation symmetry axes [1]. The feathers of goose's left wing are typically used to make shuttlecocks because they are more symmetrical, have a more consistent shape and are more flexible than those from the right wing. Shuttlecocks do not exhibit mirror symmetry and are chiral bodies, due to the way the feathers are placed into the cork, which is responsible for the anti-clockwise spinning of these projectiles as they propagate through the air. The symmetry of the badminton court is different. It corresponds to the mm2 point group. Figure 2 shows mirror symmetry at the net and 2-fold symmetry at the centre of the court. When two right-handed players (or two left-handed players) face each other (Figures 2a & 2b), there is a global 2-fold symmetry on the court, which is also contained in the symmetry of the shuttlecock. The situation is therefore symmetry-equivalent for both right-handed (or left-handed) players facing each other. When a left-handed player faces a right-handed player (Figures 2c-2d) the 2-fold symmetry is lost and a mirror symmetry appears. Since there is no mirror symmetry for the shuttlecock, the situation is not symmetry-equivalent for the left-handed and right-handed players, due to the fact that the racket manipulated by both players can induce different shuttlecock spinning by symmetry.

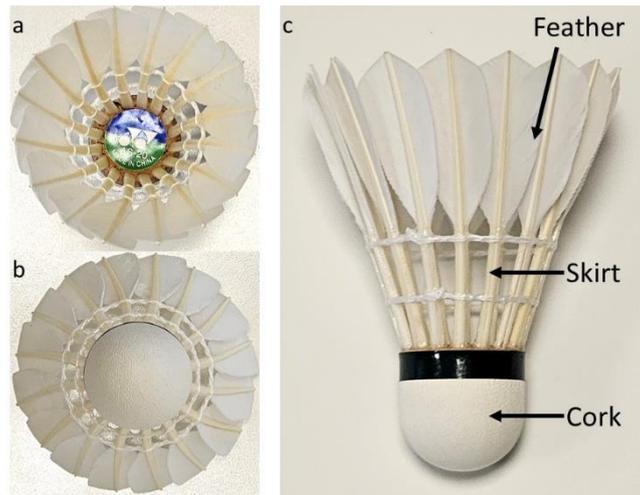

Figure 1. Structure and symmetry of a feathered shuttlecock. The top and bottom views along the rotation axis (a and b) show the 16-fold symmetry. This is due to the way the 16 bent feathers are placed into the cork (c). The shuttlecock also exhibits 2-fold symmetry.

Most of the studies of shuttlecock trajectories found in the literature [2-9] concern experimental measurements, including wind tunnel experimental data of fast camera, and/or theoretical descriptions of net shots, serve shots, smashes and high clear shots. These studies aimed at gaining knowledge in the equations of motion of a shuttlecock's flying trajectory. Usually the trajectory is described in a plane, defined by the initial position and the horizontal and vertical components of velocities. The reproducibility and quality of such studies is essential for quantitative analysis of the flying properties of the different shuttlecocks. Since the rotation axis of the shuttlecock is collinear with its velocity, it is often considered that Magnus effect [10] never occur in Badminton, contrary to other sports ball such as baseballs, volleyball and soccer [11]. Measurements of the lateral force on the shuttlecock suggested that the spinning of the shuttlecock does induce a Magnus effect, which may be the origin of the shuttlecock "drifting" noticed by players for feather shuttlecock during high clear [3, 12]. In many studies trajectories are not induced by players because the initial trajectory of the projectile after real impact with the racket is sometime hard to control or to reproduce precisely. It is also player dependent, which precludes quantitative analysis.

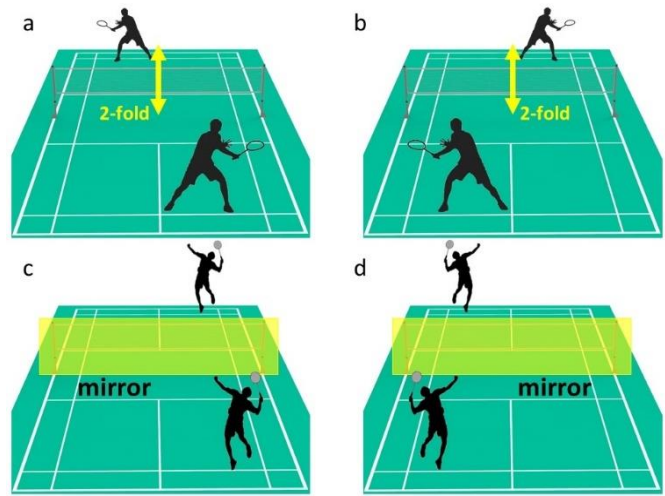

Figure 2. Sketch of a badminton court, exhibiting 2-fold symmetry around the court centre and mirror symmetry centred on the net. A situation where right-handed (a) or left-handed (b) players face each other corresponds to 2-fold symmetry. A situation where a left-hander and a right-hander face each other (c,d) corresponds to a mirror symmetry.

In this study, we present a qualitative comparison of the shuttlecock trajectories after slice shots performed by right-handed and left-handed players. Badminton slicing technique is known for causing the shuttle to spin and altering the angle of return. Here we focussed our attention on cross-court slice drop shots (Figure 3). Righties played the slice in the right-hand corner of the court, while lefties played the slice in the left-hand corner of the court. A high-speed Phantom Miro 3a10 camera was used on the badminton court, in order to capture high frame rate (3,700 fps) videos of the trajectories of the shuttlecock in real condition. At this frame rate, the resolution of the videos was maximized to 1024×768 pixels and the exposure time for each frame was set to 268 µs. We used Yonex AS 20 shuttlecocks for the slice shots. The videos are shown in supplementary material. Due to the fact that videos were collected on the court (3D), it was not possible to extract precise measurements of distances (and therefore speed) from the videos. We estimated roughly speed by scaling lengths with shuttlecock (9 cm) or racket sizes (68 cm). Given the expected readership, the choice is made to use popular units for speed (km/h) and rotational speed (rps) to make this article readable by non-physicists. In the following we discuss *i)* the shuttlecock flipping after smash *ii)* the anticlockwise spinning for right-handed slice shots and clockwise-to-anticlockwise spinning for left-handed slice shots *iii)* the shuttlecock flipping and Magnus deviation after left-handed and right-handed slice shots.

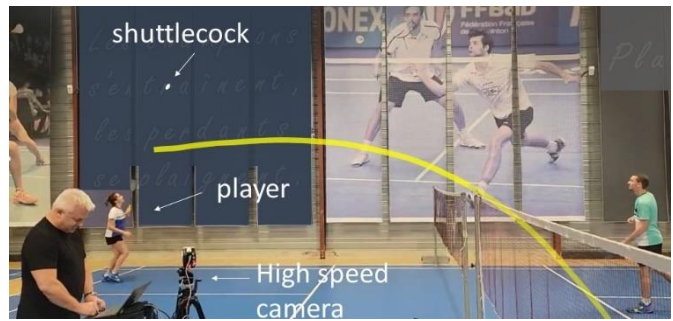

Figure 3. A high-speed camera captured videos of the trajectories of the shuttlecock at 3,700 fps on the badminton court, as players performed different cross-court slice shots.

## 2. Shuttlecock flipping after smash

It is well-known that, due to their peculiar chiral structure, badminton shuttlecocks spin as they fly in a high-drag. This specificity compared to sport ball has been subject of an intensive research from the point of view of aerodynamics [2-9], with unexpected applications such as improvements in landing accuracy of re-entry vehicles [13]. The value of the ballistic coefficient of badminton shuttlecocks is small, which translates their ability to overcome air resistance in flight [2]. The shuttlecock generates large aerodynamic drag, so that initial velocities of shuttlecocks of $\simeq$240 km/h are reduced in only $\simeq$0.6 s to near the terminal velocity of $\simeq$25 km/h [12, 14]. Women's record for fastest badminton shot is currently set to 438 km/h and belongs to P. Tan, while the men's record at 565 km/h belongs to S. Rankireddy. Both are right-handed players. Since shuttlecocks are light and extended particles, they fly with a pure drag trajectory. Kitta et al. [2] compared trajectories of spinning and non-spinning feather shuttlecock and concluded that spinning shuttlecock exhibit a marginally larger drag.

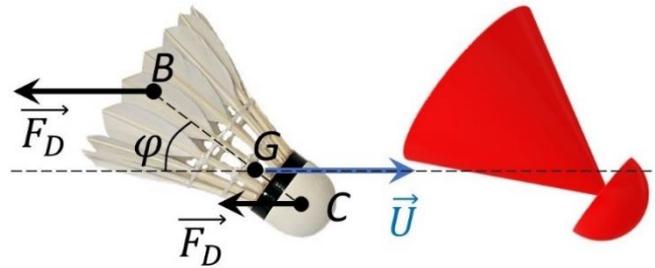

Figure 4. Left: the drag forces $F_D$ applied on the feather and the cork generate a torque, stabilizing the shuttlecock axis along its velocity (see ref [5]). Right: schematic representation used in Figure 6.

Cohen *et al* considered the effect of the aerodynamic pressure forces on the shuttlecock, with a global drag force applied at the pressure centre point at the level of the feather [5], which depends on the surrounding pressure profile. When the pressure is uniform on the feather, the aerodynamic centre is the centroid of the object. However, the centre of gravity, closer to the cork, differs from the centre of pressure [15]. Cohen *et al* introduced a model explaining the shuttlecock flipping after impact with the racket, changing the direction of the shuttlecock propagation. They have shown that the stabilizing aerodynamic torque sets the shuttlecock nose ahead after impact. They modelled the shuttlecock with two spheres (figure 4): one representing the light skirt with large cross-section (positioned in B) and one representing the heavy small cork with smaller cross-section (placed in C) [5]. They explained that when the projectile flies with an angle $\varphi$ with respect to the velocity $\vec{U}$, after an impact with a racket, a torque balance around the centre of gravity $G$, due to different drag forces on the skirt and cork, makes the shuttlecock axis collinear to velocity $\vec{U}$. For an initial shuttlecock velocity of $\simeq 10$ m.s$^{-1}$ (36 km/h), the found a flipping time of $\simeq 20$ ms, while the stabilization of the symmetry axis of shuttlecock along the direction of its velocity occurs typically within 150 ms and over $\simeq 1$ m [5, 16]. We observed such flipping processes on video 1, where the initial velocity after racket impact is $\simeq 50$ m.s$^{-1}$ (290 km/h). The flipping time is of the order of 5 ms, while the stabilization of the symmetry axis of shuttlecock occurs within $\simeq 35$ ms and over $\simeq 1$ m (one racket length is about 68 cm). We can also observe an important deformation of the shuttlecock during impact, as part of the energy is transferred to internal degrees of freedom of the shuttlecock, acting as energy bumper. Indeed, its important deceleration during impact bends the feather and the shuttlecock. We can see this torsion mode on video1 during few milliseconds, which corresponds to a breathing mode. This symmetric deformation mode of the shuttlecock is similar to jellyfish swimming or to the breathing mode of some molecules, revealed by femtosecond pump-probe techniques [17-19].

**3. Right-handed vs left-handed slices**

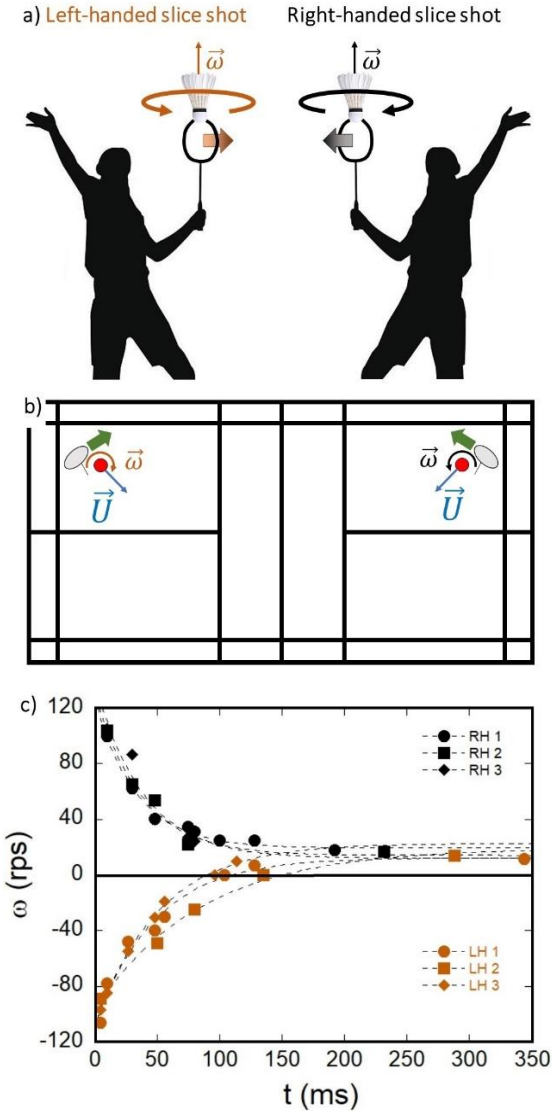

Figure 5. (a) Slice shot performed by right-handed and left-handed players, inducing opposite shuttlecock spinning after contact with racket. (b) Rotation speed $\omega(t)$ after slice shots.

Videos 2 and 3 show different slices performed by elite right-handed and left-handed players. We extracted from the video analysis the temporal evolution of the shuttlecock rotation speed $\omega(t)$ around its symmetry axis. The cross-court slice shots were performed as shown in Figures 3, 5a and 5b, with left-handed players at the left corner of the court, sending the shuttlecock in diagonal at the opposite side just behind the net. We used the mirror symmetry situation with the right-handed players performing the slice shots from the right corner of the court towards opposite side behind the net. Figure 5c shows the evolution of shuttlecock rotation speed $\omega$ obtained from 3 slice shots (RH1, RH2 and RH3) performed by right-handed players. The slice allows to generate very high spinning, of the order of 100 revolution per second ($\omega \simeq 100$ rps) few ms after contact with racket. The anticlockwise rotation is chosen as positive here, which corresponds to the natural rotation of the shuttlecock as it flies in the air. Figure 5c shows that the rotation speed rapidly decreases, with an apparent exponential decay. After 300 ms, on approaching the net, the rotation speed is of the order of 15 rps. The slice shot of left-handed players (LH1, LH2 and LH3) generates similar fast spinning ($\simeq$100 rps) just after contact with racket. However, the main difference is that the shuttlecock spins clockwise ($\omega < 0$)! As the rotation accelerates (decelerates in absolute value), the shuttlecock stops spinning after $\simeq$100 ms, then starts to spin anticlockwise due to air pressure on the chiral feather. 300 ms after contact with racket, the shuttlecock reaches $\simeq$15 rps, which is similar to the spinning observed for right-handed players. Since experimental data show an apparent exponential decay of rotation speed, we used the following empirical model to fit the curves:

$$\omega(t) = \omega_i \exp(-\frac{t}{\tau}) + \omega_f \qquad (1).$$

$\omega_i$ corresponds approximately to the initial rotation speed, just after contact with the racket, $\omega_f$ to the rotation speed after 300 ms and $\tau$ is the deceleration time constant, which depends on shuttlecock, air density... We used independent fits for the different slice shots (RH1, RH2, RH3, LH1, LH2 and LH3) and the results are given in table 1. The average initial rotation just after contact with the racket are similar for right-handed and left-handed players ($\simeq$120 rps). The deceleration time constants are similar for the 3 right-handed players shots (≃36 (9) ms) and globally larger for left-handed players (61(10) ms), while final rotation speeds on approaching the net after 300 ms are similar (≃15 rps). These very fast changings in rotation speeds can be clearly observed in video 4, slowdown 500 times. As slice shots are performed, players and coaches noticed different sounds for left-handed and right-handed player. The is due to the orientation of the feathers: the racket of the right-handed player rubs the shuttlecock almost parallel to the feathers, while it is almost perpendicular for the left-handed players. It may therefore be easier for left-handed players to spin the shuttlecock, but this process is too fast to be studied with the present videos set-up.

|        | $\omega_i$ | $\tau$   | $\omega_f$ |
|--------|-----------|----------|-----------|
| RH1    | 107 (6)   | 33 (3)   | 19 (2)    |
| RH2    | 118 (13)  | 38 (12)  | 14 (7)    |
| RH3    | 127 (20)  | 36 (9)   | 12 (11)   |
| <RH>   | 118 (17)  | 51 (4)   | 15 (9)    |
| LF1    | −131 (3)  | 84 (11)  | 22(4)     |
| LF2    | −116 (5)  | 48 (9)   | 19(6)     |
| LF3    | −121 (9)  | 48 (9)   | 12 (7)    |
| <LF>   | −123 (9)  | 61 (10)  | 16 (7)    |

Table 1. Fitting parameters from equation (1) for the different slice shots for right-handed and left-handed players.

The high rotational speed of the shuttlecock is generating a large centrifugal force. Video 5 shows the deformation induced by fast spinning perpendicular to velocity. The drag coefficient for a rotating standard shuttlecock can increase due to a large centrifugal force [20]. Obayashi *et al* investigated the effect of a shuttlecock rotation on its skirt deflection [2]. They have shown that the skirt enlargement due to the centrifugal forces is compensated by the effect of the aerodynamic drag, when the shuttlecock travels aligned along its rotation axis. We can see on the different videos that there is no significant shape change when shuttlecock rotation axis is along velocity, while shuttlecock feathers clearly bend when the spinning axis is perpendicular to velocity, as aerodynamic drag does not compensate deformation induced centrifugal force.

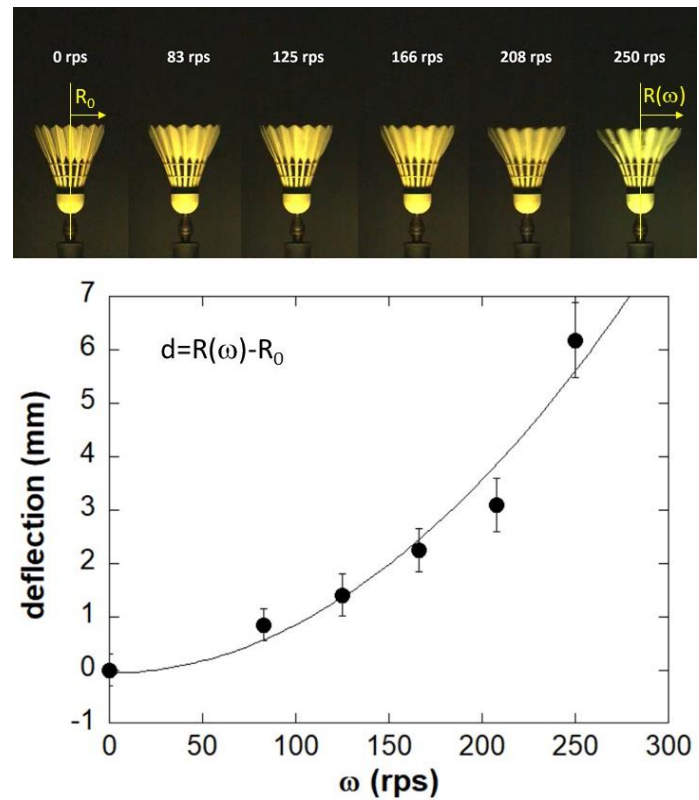

Figure 6. Shuttlecock deformation with rotation speed (top). The radius deflection *d* almost scales with $\omega^2$, as shown by the fit of the data.

Figure 6 and Video 5 show the deformation of the shuttlecock, mounted on a Dremel, with rotation speed. The deformation of the shuttlecock is characterized by the deflection *d* of the radius $R(\omega)$ measured at rotation speed $\omega$, with respect to the radius at rest $R_0$ with $d = R(\omega) - R_0$. This deflection is due to the high centrifugal force, related to centripetal acceleration $a_c$, acting on the feather. Considering $a_c = R\omega^2$, with $\omega$=785 rad.s$^{-1}$ (125 rps) and $R$=0.0335 m, the centripetal acceleration at the edge of the feather reaches 20 643 m.s$^{-2}$, which corresponds 2 000 g. Even if the mass of the feather is small, the very high g force acting on its mass is able to bend it. If we approximate the mass density on the feather as homogeneous along its lengths, the deflection is then simply proportional to the g force, itself scaling with $\omega^2$. The fit in Figure 6, considering $d \propto \omega^2$ is in quite good agreement with experimental data, given the rough approximation considering the feather as homogeneous. The shuttlecock deformation observed on video 5 just after the slice is similar to the one observed around 125-166 rps in Figure 6, which confirms that its deformation is due to its high spinning rate just after slice shot.

The clockwise to anticlockwise shuttlecock rotation induced by the left-handed players slice shot is therefore different from the one of the right-handed players, as suspected with the symmetry considerations introduced above, and clearly shown by the super-slow-motion videos. For right-handed players, the faster spinning may accelerate the shuttlecock, while for left-handed players the anticlockwise rotation should slow down the shuttlecock. We will discuss trajectories in more details later, because the reader may have noticed a strange phenomenon on videos 2 and 3.

**4. Magnus affect after slice shot**

Videos 2 and 3 show that, for some slice shots, after contact with the racket the shuttlecock is no more oriented along its initial speed $\vec{U}$, but almost perpendicular. For righties (Figure 7a), the shuttlecock points towards the player. Here again the initial velocity towards the net is associated with drag forces (Figure 4) responsible for a torque stabilizing the shuttlecock with the cork pointing towards its velocity. Videos 2 and 4 and Figure 7a clearly show that in addition to flipping towards its stable orientation, the shuttlecock is lifting up during the early stage of the fly. This phenomenon is due to the Magnus effect [10]. When the shuttlecock rotates perpendicularly to its velocity as it flies, the surrounding air rotates. Therefore, the air speed on one side increases and decreases on the other side. Bernoulli's theorem indicates that an increase in fluid velocity leads to a decrease in pressure, which results in a transverse force, the Magnus Force ($\vec{F_M}$), acting on the shuttlecock. We mentioned above that Magnus effect is usually neglected in badminton, because the rotation axis is usually colinear with velocity. However, the videos shown here clearly reveal that after impact with the racket the rotation axis can be almost perpendicular to the velocity during the first 30 ms, when the rotation speed $\vec{\omega}$ is very high (above 50 rps, Figure 5c).

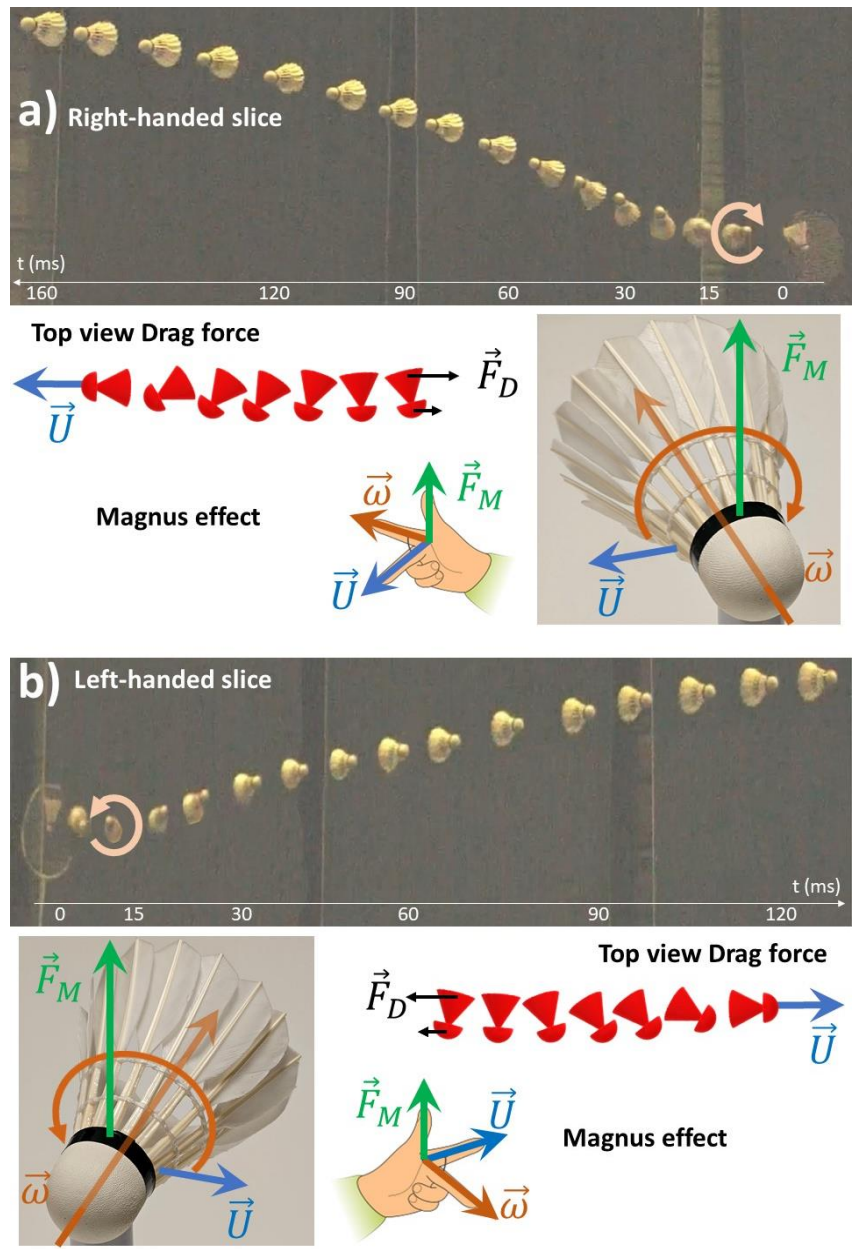

Figure 7. Magnus effect lifting-up the projectile after slice shots performed by right-handed (a) or left-handed player (b). It occurs in the early stage of the trajectory, as the anti-clockwise (righties) or clockwise (lefties) spinning axis $\vec{\omega}$ of the shuttlecock is perpendicular to its velocity $\vec{U}$. The Magnus effect stops when spinning axis gets collinear to velocity, due to the drag torque.

The Magnus force is given by $\overrightarrow{F_M} \propto \rho V \vec{\omega} \wedge \vec{U}$, where ρ is the air density and V the volume of shuttlecock. The angular right-hand rule is very useful to explain the Magnus effect from a right-handed player (Figure 7a): the right index points in the direction of the rotation ($\vec{\omega}$), the finger index points in the direction of the velocity ($\vec{U}$) and the thumb is then pointing towards the Magnus force ($\overrightarrow{F_M}$). The Magnus force is then lifting up the shuttlecock. This effect stops within ≃30 ms as the shuttlecock stabilizes his spinning axis $\vec{\omega}$ collinear to its velocity $\vec{U}$. It is therefore necessary to use super-slow-motion cameras to observe this transient Magnus effect and the fast-initial spinning responsible for this phenomenon. If we look at the trajectory of the shuttlecock after a left-handed player slice shot (Figure 7b), we can see that it also lifts up. Indeed, the left-handed player spins the shuttlecock in opposite directions, compared to the right-handed player, but the initial velocity also changes direction (Figure 5b). The cross product is unchanged and the Magnus force also lifts up the shuttlecock. For the 10% left-handed readers, in addition to the right-handed rule mentioned above, it is suggested to use for this specific case the angular left-hand rule (Figure 7b) to understand the direction of the Magnus effect: the left index points in the direction of the velocity ($\vec{U}$), the finger index in the direction of the rotation ($\vec{\omega}$) and the thumb is pointing towards the Magnus force ($\overrightarrow{F_M}$), which is also lifting-up the shuttlecock. The amplitudes, timescales and trajectories of the shuttlecock lifting during the Magnus effect are similar for right-handed and left-handed players. This is due to the fact that during both types of slice shots the initial rotation speeds and velocities are similar for lefties and righties. The magnitude of the Magnus effect observed in videos 2 and 3 is in the range of 20-70 cm. When the player gives a large transverse component to the initial velocity of the shuttlecock (with titled racket or shot), the shuttlecock flips and its rotation axis $\vec{\omega}$ gets perpendicular to its velocity $\vec{U}$, maximizing so the cross product behind the Magnus force lifting up the shuttlecock. As this transverse component is reduced, the Magnus effect lowers in amplitude. When the shuttlecock spinning axis is perpendicular to its velocity, the higher the initial spinning and velocities, the stronger the Magnus effect. The Magnus effect can therefore modify significantly the trajectories for both left-handed and right-handed players.

## 5. Shuttlecock trajectories

Video 6 and Figure 8 show the global trajectories for typical slice shots performed by left-handed and right-handed players. For slice-shot crossing the court over a similar distance, the time of flight is about 20% longer for left-handed player slice shots (0.91 (4) s vs 1.08 (4) s). This is due to the lowest average horizontal speed as the shuttlecock is slow-down by its clockwise initial spinning. As the shuttlecock reaches the net, the lowest horizontal speed translates in a more vertical trajectory, approaching a parabola. The falling angle of the trajectory as the shuttlecock touches the ground is around $\simeq 65°$ for a left-handed player compared to $\simeq 50°$ for a right-handed player. Consequently, if the shuttlecock passes just above the net for both types of players, it will fall closer to the net in the case of a slice shot performed by a left-handed player, compared to a right-handed player. The different ratios between horizontal and vertical speeds is therefore due to the clockwise spinning of the shuttlecock after slice shot performed by left-handed players, which is rotating in opposite direction to the natural spinning. Previous aerodynamic studies of spinning shuttlecock revealed a weak variation of the drag force with or without spinning, which may explain the slower trajectory for left-handed players. However, in the case of the clockwise spinning, opposite to natural rotation, the drag force and the aerodynamics may change significantly.

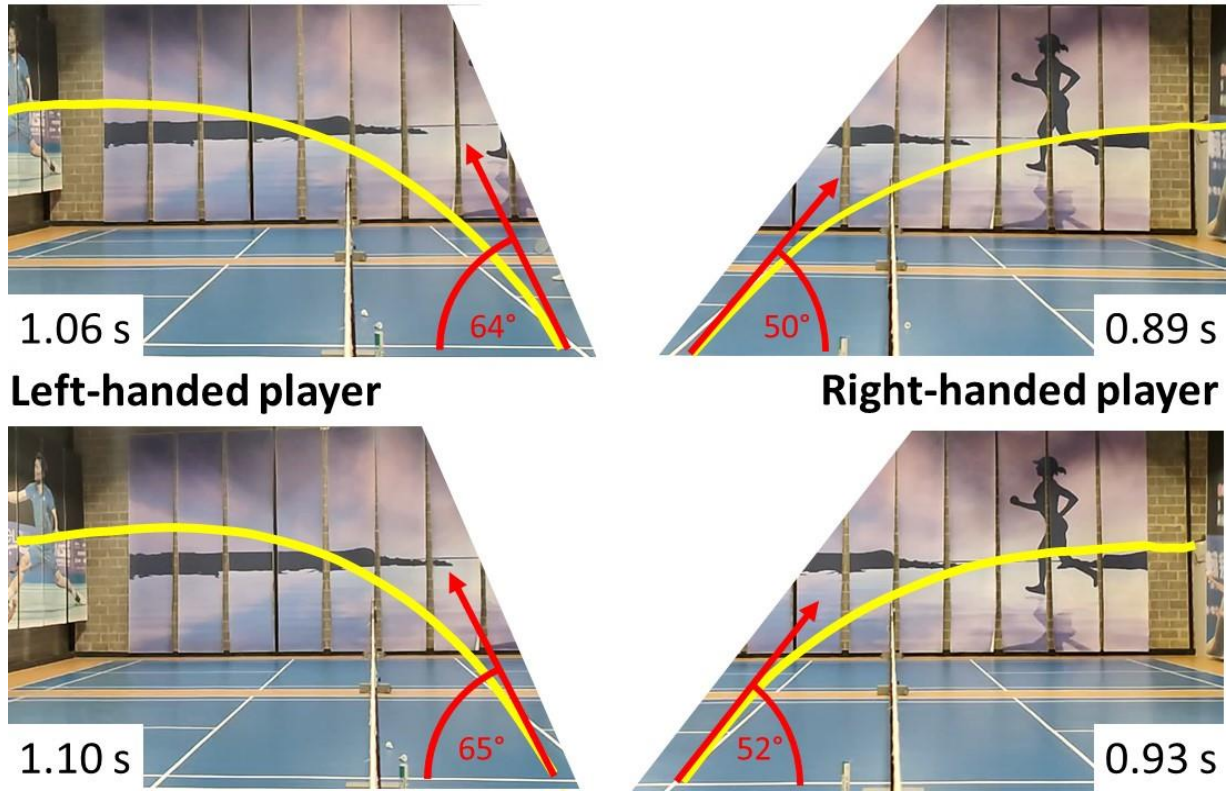

Figure 8. Typical trajectories for slice shots (yellow) performed by left-handed (left) and right-handed (right) players. The time of flight is longer for left-handed players and the slower trajectory translate in a higher falling angle (red) for lefties (≃65°) than righties (≃51°).

## 6. Conclusion

Super-slow-motion videos provide key information on ultrafast dynamics intrinsic to badminton physics, driven by extreme stresses applied to the shuttlecock after racket impact. For manufacturers, it is a real challenge to propose feather shuttlecocks able to support ultimate spinning rates (larger than 100 rps), ultrafast flipping or deformation breathing after impact. Indeed, there is often a moment in the game where players have to perform strategic powerful shots or fast slices. The videos revealed very different spinning dynamics, caught on the fly, after slice shots performed by lefties and righties. The accelerated counter clockwise spinning induced by right-handed players and the clockwise to anti-clockwise spinning induced by left-handed players explains the different trajectories of the shuttlecock, creating a steeper sliced shot for left-handed players. Although how much of an advantage this gives lefties is not clear, it is an advantage. Hopefully, this is somewhat compensated by the possibility for the players to perform reverse-slice shots, which are then more advantageous for righties and which could be a subject for future studies. It is important to underline that a slice shot is efficient when it skims the net. The present results highlight the importance of balancing the transfer of transverse speed to the shuttlecock, to limit the Magnus effect. In fact, too much Magnus effect raises the shuttlecock more than necessary and thus reduces the effectiveness of the slice shot. This concerns both right-handed and left-handed players. The players are thus left to make the right shot and as Robbin Williams said "What's right is what's left if you do everything else wrong".

**Supplementary material**

The videos are available on a YouTube channel: click the links:

**Video 1:** Shuttlecock flipping and breathing. https://youtu.be/w2xsVHsD8J8

**Video 2a & 2b:** slice right-handed players https://youtu.be/UgeZOBo--5M & https://youtu.be/uxcP8iKRBVc

**Video 3a & 3b:** slice left-handed players https://youtu.be/PivDQ0f4h5k & https://youtu.be/edMxKmq-E4M

**Video 4:** time evolution of rotation speed https://youtu.be/9KG0-MBja6s

**Video 5:** centrifugal force on the shuttlecock https://youtu.be/jNNvOnucV0Q1001

**Video 6:** shuttlecock trajectories https://youtu.be/5HjanVzecrE

The videos are available on Zenodo with doi 10.5281/zenodo.10012876

**Acknowledgements**

E. Collet would like to thank the players who participated in this study for their support and patience (F. Noël and L. Dufay, C. Dubois, G. Bodin, H. Crespel, N. Latimier, R. Cloteaux-Foucault and M. Justel), the coaches for their interesting comments (M. Dureault, J. Bloutin, P. Chagnot, E. Kehlhoffner and R. Bridel) as well as the Flume Ille Badminton club and the Plus2Bad arena for their help for the video shooting. E. Collet would like also to thank A. Gauthier, I. Cantat and J.F. Le Guennan for their technical support for the use of the fast camera.

**Déclaration d'intérêts**

L'auteur ne travaille pas, ne conseille pas, ne possède pas de parts, ne reçoit pas de fonds d'une organisation qui pourrait tirer profit de cet article, et n'a déclaré aucune autre affiliation que ses organismes de recherche.

**Financement**

Ce travail a été soutenu par l'Institut Universitaire de France et l'Agence Nationale de la Recherche, ANR-19-CE30-0004 ELECTROPHONE.

# References

[1] IUCr, International Tables for Crystallography, Volume A: Space Group Symmetry, 5. revised edition ed., Kluwer Academic Publishers, Dordrecht, Boston, London, 2002.
[2] S. Kitta, H. Hasegawa, M. Murakami, S. Obayashi, Aerodynamic properties of a shuttlecock with spin at high Reynolds number, Procedia Engineering, 13 (2011) 271-277. https://doi.org/10.1016/j.proeng.2011.05.084
[3] C.M. Chan, J.S. Rossmann, Badminton shuttlecock aerodynamics: synthesizing experiment and theory, Sports Engineering, 15 (2012) 61-71. 10.1007/s12283-012-0086-7
[4] A. Verma, A. Desai, S. Mittal, Aerodynamics of badminton shuttlecocks, Journal of Fluids and Structures, 41 (2013) 89-98. 10.1016/j.jfluidstructs.2013.01.009
[5] C. Cohen, B.D. Texier, D. Quéré, C. Clanet, The physics of badminton, New Journal of Physics, 17 (2015) 063001. 10.1088/1367-2630/17/6/063001
[6] R.C. Mehta, Aerodynamics of Sport Balls, Badminton Shuttlecock and Javelin, Scholars Journal of Engineering and Technology, 11 (2023) 4-10. 10.36347/sjet.2023.v11i02.001
[7] L.M. Chen, Y.-J. Pan Yh Fau - Chen, Y.J. Chen, A Study of Shuttlecock's Trajectory in Badminton.
[8] B.D. Texier, C. Cohen, D. Quéré, C. Claneta, Shuttlecock dynamics, Procedia Engineering, 34 (2012) 176-181. https://doi.org/10.1016/j.proeng.2012.04.031
[9] J.E. Goff, A review of recent research into aerodynamics of sport projectiles, Sports Engineering, 16 (2013) 137-154. 10.1007/s12283-013-0117-z
[10] G. Magnus, Ueber die Abweichung der Geschosse, und: Ueber eine auffallende Erscheinung bei rotirenden Körpern, Ann Phys-Berlin, 164 (1853) 1-29. https://doi.org/10.1002/andp.18531640102
[11] B.D. Texier, C. Cohen, D. Quéré, C. Clanet, Physics of knuckleballs, New Journal of Physics, 18 (2016) 073027. 10.1088/1367-2630/18/7/073027
[12] A.J. Cooke, Shuttlecock aerodynamics, Sports Engineering, 2 (1999) 85-96. https://doi.org/10.1046/j.1460-2687.1999.00023.x
[13] J. Nicholas, C. Eun Jung, 2D aerodynamic analysis of a badminton shuttle for re-entry vehicle applications, Proc.SPIE, 2020, pp. 1137615.10.1117/12.2558935
[14] M. Hubbard, Spin dynamics of the badminton shuttlecock, 1997,
[15] A. Cooke, Computer simulation of shuttlecock trajectories, Sports Engineering, 5 (2002) 93-105. https://doi.org/10.1046/j.1460-2687.2002.00100.x
[16] V. Kaushik, Unique Flight Features of Shuttlecock, International Research Journal of Engineering and Technology, 4 (2017) 2395. 10.1038/ncomms12212
[17] H.T. Lemke, K.S. Kjær, R. Hartsock, T. Brandt van Driel, M. Chollet, J.M. Glownia, S. Song, D. Zhu, E. Pace, S.F. Matar, M.N. Nielsen, M. Benfatto, K.J. Gaffney, E. Collet, M. Cammarata, Coherent structural trapping through wave packet dispersion during photoinduced spin state switching, Nat. Commun., 8 (2017) 15342.
[18] S. Zerdane, L. Wilbraham, M. Cammarata, O. Iasco, E. Rivière, M.L. Boillot, I. Ciofini, E. Collet, Comparison of structural dynamics and coherence of d–d and MLCT light-induced spin state trapping, Chem. Sci., 8 (2017) 4978-4986. 10.1039/c6sc05624e
[19] M. Cammarata, R. Bertoni, M. Lorenc, H. Cailleau, S. Di Matteo, C. Mauriac, S.F. Matar, H. Lemke, M. Chollet, S. Ravy, C. Laulhe, J.F. Létard, E. Collet, Sequential Activation of Molecular Breathing and Bending during Spin-Crossover Photoswitching Revealed by Femtosecond Optical and X-Ray Absorption Spectroscopy, Physical Review Letters, 113 (2014) 227402. Artn 227402

Doi 10.1103/Physrevlett.113.227402
[20] H. Hasegawa, S. Kitta, M. Murakami, S. Obayashi, Flow analysis and aerodynamic characteristics of a badminton shuttlecock with spin at high Reynolds numbers, Sports Engineering, 16 (2013) 91-98. 10.1007/s12283-013-0112-4